# Sex chromosome evolution:
# The classical paradigm and so much beyond

Paris Veltsos[1,2], Sagar Shinde[1], Wen-Juan Ma*[1,3]

orcid.org/0000-0002-8872-6281 Paris Veltsos

orcid.org/0000-0003-2585-6406 Wen-Juan Ma

1. Research group of Ecology, Evolution and Genetics, Biology Department, Vrije Universiteit Brussel, Brussels, Belgium
2. Department of Ecology and Evolutionary Biology, University of Kansas, Kansas, USA
3. Department of Molecular Biosciences, University of Kansas, Kansas, USA

*Correspondence:

Wen-Juan Ma: wen-juan.ma@vub.be, 0032 2629 3416

# Abstract

Sex chromosomes have independently evolved in species with separate sexes in most lineages across the tree of life. However, the well-accepted canonical model of sex chromosome evolution is not universally supported. There is no single trajectory for sex chromosome formation and evolution across the tree of life, suggesting the underlying mechanisms and evolutionary forces are diverse and lineage specific. We review the diversity of sex chromosome systems, describe the canonical model of sex chromosome evolution, and summarize studies challenging various aspects of this model. They include evidence that many lineages experience frequent sex chromosome turnovers or maintain homomorphic sex chromosomes over long periods of time, suggesting sex chromosome degeneration is not inevitable. Sometimes the sex-limited Y/W chromosomes expand before they contract in size. Both transposable elements and gene gains could contribute to this size expansion, which further challenges gene loss being the hallmark of sex chromosome degeneration. Finally, empirical support for the role of sexually antagonistic selection as a driver of recombination suppression on sex chromosomes remains elusive. We summarize models that result in loss of recombination without invoking sexually antagonistic selection, which have not been empirically verified yet, and suggest future avenues for sex chromosome research.

# Glossary

**Achiasmy**: Meiosis without chiasma formation in one or both sexes.
**Androdioecy**: Breeding system in which males and hermaphrodites coexist within a population.
**Dosage compensation**: The equalization of gene expression between sexes, to accommodate for loss of gene function on the sex-limited chromosomes.
**Evolutionary strata**: Distinct regions with punctuated sequence divergence, interpreted as regions in which recombination stopped between the heterologous sex chromosomes at different times.
**Female heterogametry**: Sex chromosome system in which females have heterologous sex chromosomes (ZW females, ZZ males).
**Gynodioecy**: Breeding system in which females and hermaphrodites coexist within a population.
**Haplodiploidy**: Sex-determining system where diploid females develop from fertilized eggs and haploid males develop from unfertilized eggs.
**Heterochiasmy**: Big difference in recombination rate or location between the two sexes.
**Heteromorphic sex chromosome**: Sex chromosomes that can be distinguished under light microscopy.
**Homomorphic sex chromosomes**: Sex chromosomes that cannot be distinguished with light microscopy. They may exhibit genetic differentiation at the sequence level.
**Male heterogametry**: Sex chromosome system in which males have heterologous sex chromosomes (XY males, XX females).
**Pseudoautosomal region (PAR)**: The region where sex chromosomes still recombine.
**Sex chromosome turnover**: The use of different chromosomes to determine sex in closely related species. They may have different sex-determining loci.
**Sexually antagonistic (SA) allele**: An allele that is beneficial to the fitness of one sex but deleterious to the other.
**Transposable elements (TEs)**: DNA sequences that can move within a genome, sometimes affecting genes and contributing to genome size variation.

# Key points

1. Sex chromosomes have independently evolved in species with separate sexes in most lineages across the tree of life. The most common sex chromosome systems include male heterogametic (XY), female heterogametic (ZW), the mating type and UV system.
2. There is no single trajectory for sex chromosome formation and evolution across the tree of life. We review the diversity of sex chromosome systems, describe the canonical model of sex chromosome evolution, and summarize studies challenging various aspects of this model.
3. The challenging aspects include many lineages experiencing frequent sex chromosomes turnovers or maintaining homomorphic sex chromosomes over long periods of time.
4. There are cases showing that the sex-limited Y/W chromosomes expand before they contract in size.
5. Both transposable elements and gene gains could contribute to this size expansion.
6. Finally, empirical support for the role of sexually antagonistic selection as a driver of recombination suppression on sex chromosomes remains elusive.
7. We summarize models that result in loss of recombination without invoking sexually antagonistic selection, which have not been empirically verified yet, and suggest future avenues for sex chromosome research.

## Introduction

### Diverse sex chromosome systems across the tree of life

The sex chromosomes are a pair of chromosomes in which the sex-determining locus is located. They have independently evolved numerous times in species with separate sexes in animals, plants, fungi etc. across the tree of life (Fraser and Heitman 2004; Ming et al. 2011; Ashman et al. 2014; Bachtrog et al. 2014; Beukeboom and Perrin 2014). The fundamental property of sex chromosomes is that they occur in each sex at different frequencies relative to the autosomes, which occur in both sexes at a 50% frequency. The sex with the heterologous (non-identical) pair of sex chromosomes is the heterogametic sex. In organisms where sex is expressed at the diploid stage, the sex chromosome system is named XY if the heterogametic sex is the males, and ZW if it is the females (Figure 1A). Diploid XY and ZW systems are the most well-known and prevalent in eukaryotes. The most well-studied XY systems are those of mammals and *Drosophila* fruit flies, while the most well-studied ZW systems include those in birds and butterflies (Ezaz et al. 2006; Ellegren 2011; Figure 1B). In both XY and ZW sex chromosome systems, the sex-limited chromosome (Y or W) is present in the heterogametic sex all the time, and their homologous X or Z chromosome spends 1/3 or 2/3 of its time in the heterogametic and homogametic sex, respectively. There are also organisms that express sex at the haploid stage such as algae, moss, fungi and yeast, and their sex chromosome system is called a UV system, or mating-type system in isogamous species without sexual dimorphism between gamete types (Fraser and Heitman 2004; McDaniel et al. 2013; Lipinska et al. 2017). In these organisms, diploid individuals have both sex/mating-type chromosomes and haploid individuals either have the V or the U chromosome, or one of the mating-type chromosomes (Figure 1A).

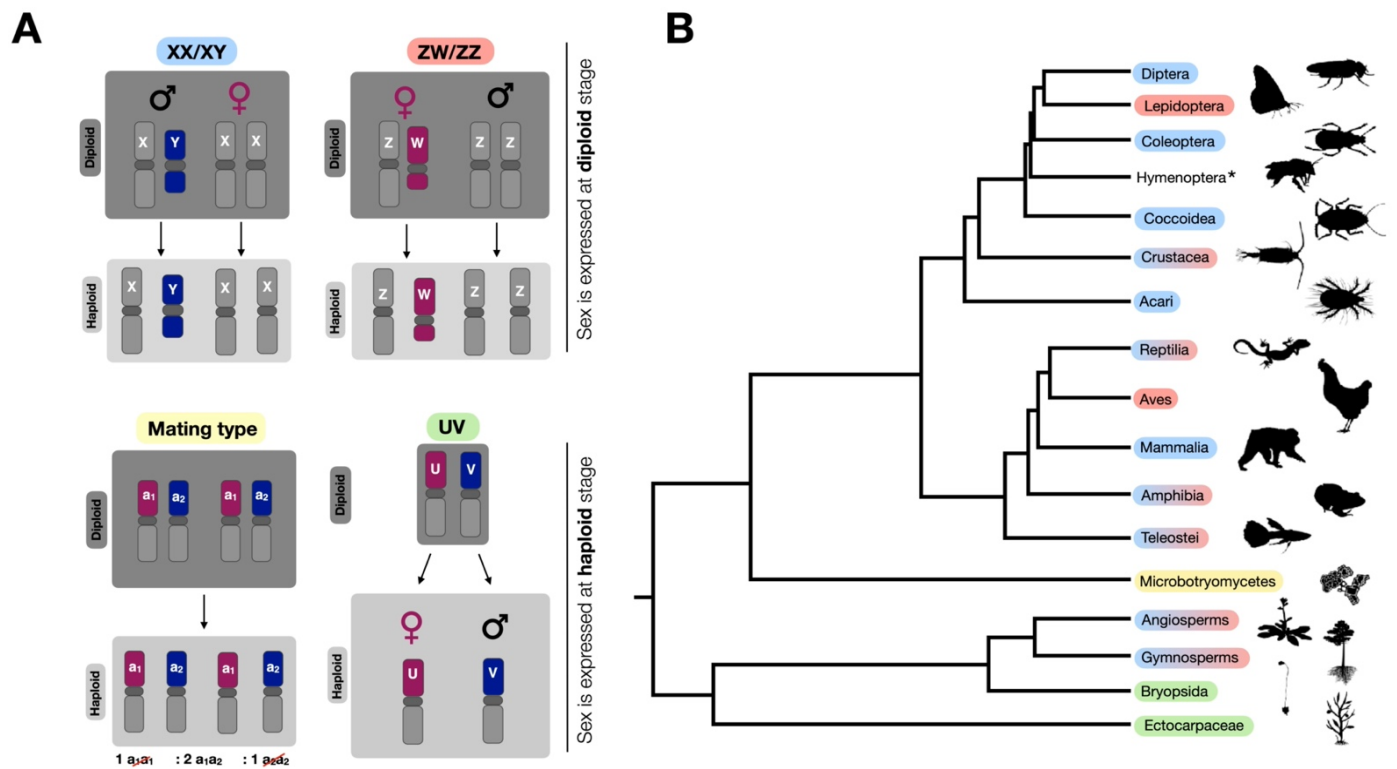

**Figure 1**. Cartoon illustration of (A) the most common sex chromosome systems and (B) their distribution in major lineages across the tree of life. A. In male heterogametic (XY) and female heterogametic (ZW) systems, sex is expressed at the diploid phase and the haploid phase is transient and short. In the mating type and UV system, sex is expressed at the haploid phase which is long in the UV system and short in the mating-type system. B. Some lineages are restricted to one of the sex chromosome systems, while others have more than one. The haplodiploid sex-determining system is indicated by an *. The phylogenetic tree was obtained from https://timetree.org, and all illustrations were obtained from https://www.phylopic.org.

Sub-variants of the XY and ZW systems involving chromosomal events such as fusion/fission and gain/loss add to the diversity of sex chromosome systems. For example, in some species the chromosome specific to the heterogametic sex is completely absent, presumably because it has been lost. This is referred to as an XO system, common in Orthoptera (White 1978), where XX are females and XO are males, or ZO system where ZO are females and ZZ are males (found in several moths) (Jaquiéry et al. 2012; Blackmon and Demuth 2014; Kageyama et al. 2017). Some organisms have multiple sex chromosomes which can arise from chromosomal fusions between the sex chromosomes and autosomes. For example, a fusion between an autosome and the Y, such as in *Drosophila miranda* (Steinemann and Steinemann 1998) and the plant *Rumex hastatulus* (Beaudry et al. 2017), generates a novel X chromosome ($X_2$). Similarly, a X-autosome fusion in a XO system, as in the grasshopper *Podisma pedestris* (John and Hewitt 1970), will generate a neo-XY system, as the partner of the autosome that fused becomes a new Y chromosome.

More sex chromosome systems exist but they are understudied, with few cases of each having been explored. For example, monotremes such as the platypus have multiple sex chromosomes which are prevented from improper meiotic segregation by reciprocal translocations between non-homologous pairs, so that non-homologous X and Y chromosomes always segregate together (Grützner et al. 2004). Certain Amazonian frogs and a Taiwanese ranid frog (both are XY system) are documented to have multiple sex chromosomes likely due to chromosomal translocations, which result in a long chain of chromosomes during male meiosis to promote correct chromosome segregation (Gazoni et al. 2018; Miura et al. 2021; Katsumi et al. 2022). Multiple sex-determining genes on different chromosomes, which differ among populations, but can be brought together by crosses, have been described in houseflies (Inoue et al. 1983; Kozielska et al. 2008). Some species such as the Japanese frog (*Glandirana rugosa*) and tropical claw frog (*Xenopus tropicalis*) (Ogata et al. 2003; Roco et al. 2015) have both XY and ZW chromosomes within the same species, and sometimes the two systems can occur in the same population such as in certain poecilid fishes (Volff and Schartl 2001), or the Australian burrowing frog *Platyplectrum ornatum* (Schimek et al. 2022). When multiple sex chromosome systems coexist in the same species or population, there often are dominant and recessive feminizing or masculinizing effects among them, and they are typically considered an unstable and transitory stage between XY and ZW systems because of the unbalanced sex ratios they generate (Beukeboom and Perrin 2014).

## The canonical model of sex chromosome evolution

### Sex-limited chromosomes are restricted to one sex

Regardless of the complexity of a particular sex chromosome system in question, sex chromosomes always have a non-random association with sex. This fundamental property of sex chromosomes led to the proposition that natural selection would specialize sex chromosomes to each sex through adaptations. For example, since individuals of the heterogametic sex are always carrying the Y or W chromosome, novel mutations that benefit the heterogametic sex arising on those chromosomes will always be exposed to (positive) selection and would tend to fix. Similarly, the X is expected to accumulate female-beneficial mutations (or Z male-beneficial mutations) if they are not fully recessive, since those chromosomes spend 2/3 of their time in XX females or ZZ males, respectively. The accumulation of genes with sex-beneficial effects on sex chromosomes should therefore be faster than on autosomes. The same principle applies to mutations with deleterious effects. For example, deleterious mutations affecting female fitness that arise on the Y chromosome (or deleterious to males on the W) are never exposed to the purifying selection that would remove them, so they are expected to accumulate on these chromosomes.

The sex chromosomes present in both sexes (X or Z) are always in single copy in the heterogametic sex. Consequently, mutations on the X or Z chromosomes affecting the heterogametic sex will be exposed to selection even if they are recessive since there is no alternative copy on the other sex chromosome (Y or W). This only applies when the Y or W chromosomes have degenerated (many functional genes are lost). In these cases, the X will become enriched in recessive male-beneficial genes and the Z in recessive female-beneficial genes. Over evolutionary time, the expectation is therefore for the Y and Z chromosomes to specialize for male function and for the X and W chromosomes to specialize for female function due to dominant or partially dominant female-beneficial genes, and for male function due to recessive male-beneficial genes.

Interestingly, the same mutations could be simultaneously beneficial for one sex but deleterious for the other, which is referred to as sexual antagonism (Charlesworth and Crow 1978). When such genes follow autosomal inheritance, they have a 50% chance of turning up in the sex in which they increase their frequency and 50% in the opposite sex, so that they on average experience balancing selection (Levene 1953). However, the sex-biased inheritance arising from sex linkage would allow such genes to increase in frequency or fix, depending on the relative benefit and disadvantage they confer to each sex. Classical models of sex chromosome evolution propose that sexually antagonistic (SA) genes could become linked to the sex-determining locus on the sex-limited Y or W chromosomes, for example, if they were brought together by a chromosomal inversion (Fisher 1931; Charlesworth and Crow 1978; William R Rice 1987a; Charlesworth 1991). A mutation bringing together the sex-determining locus and a SA gene is sufficient to generate a new sex chromosome that is selected to fix in a population.

## **Sexual system and the evolution of sex chromosomes**

Regardless of the ancestral state from which sex chromosomes (and dioecy) evolved, they originate from a pair of autosomes. The ancestral state in flowering plants is hermaphroditism, in which both sexual functions coexist in the same individual (Charlesworth 2002). Alternatively, the ancestral state in many fishes and reptiles is environmentally-dependent sex determination, in which certain environmental factors, such as temperature, can trigger pre-existing sex determination pathways towards male or female development (Baroiller et al. 2009; Georges and Holleley 2018). The evolution of sex chromosomes in this case results from a mutation that overrides the environmental factors and takes over the activation of the male or female developmental pathway (Charlesworth 1996; Muralidhar and Veller 2018). In the case of ancestral hermaphroditism, the sex-determining gene could be a mutation that knocks out one of the sexual functions to result in a mixed population with hermaphrodites and either males (androdioecy) or females (gynodioecy). As the individuals with one sexual function further evolve to specialize in it, the hermaphrodite individuals respond by specializing in the opposite sexual function until they only serve one sexual function and the population has become dioecious (Bawa 1980; Charlesworth 2002; Ming et al. 2011).

In both ancestral hermaphroditism and environmentally dependent sex-determining system, SA genes can aid the establishment of newly formed sex chromosomes, because they increase the fitness of individuals with the specialized male and female functions, which then outcompete the individuals with the ancestral sex-determining system. In addition, the replacement of environmental-dependent sex determination by genetic sex determination is thought to be beneficial, because it ensures a 50% sex ratio, especially in unstable and unpredictable conditions in which the environmental-dependent sex-determining system can result in biased sex ratios, and in organisms with shorter generation times which are most heavily impacted by sex ratio selection (Grossen et al. 2011; Quinn et al. 2011; Beukeboom and Perrin 2014).

## **The crucial role of SA genes**

The next step after the establishment of the sex-determining region is the establishment of close linkage with a SA gene. The resulting sex chromosomes are called neo (new) or proto (first) sex chromosomes. The sex-determining region is generated by a single mutation (e.g. most animals) or is itself a small non-recombining region in the two-mutation model involving hermaphrodite ancestors (e.g. flowering plants) (William R Rice 1987a; Charlesworth 1991). An example of a SA gene is (are) the gene(s) encoding high pigmentation in guppies. Pigmentation is beneficial to males because it is a sexually selected trait, but it is deleterious to females because it increases predation risk (Endler 1980). A mutation that restricts recombination between the SA gene and the sex-determining locus will be beneficial, because a Y chromosome that does not recombine will have more offspring with high fitness than its ancestor in which the SD and SA loci do recombine. The typical mutation invoked to reduce recombination is a chromosomal inversion in the canonical model; however, other mutations such as those that affect methylation status, centromere expansion or transposable element activity can have the same effect (Wang et al. 2022).

## **Sex chromosome degeneration and evolutionary strata**

Repeated cycles of novel SA mutations arising and fixing close to the non-recombining region followed by mutations that extend the region of reduced recombination are hypothesized to characterize the early evolution of sex chromosomes. The expected outcome is the formation of sex chromosome regions that stopped recombining at different times. Lack of recombination results in the eventual degeneration of the sex-limited Y or W chromosome, due to a combination of evolutionary forces such as Muller's ratchet, genetic hitchhiking and background selection (Charlesworth and Charlesworth 2000; Bachtrog 2008; Wright et al. 2016). Muller's rachet is a consequence of genetic drift, where any genetic variation that is lost by chance cannot be reintroduced by recombination. Such losses are exacerbated by the small effective population size of the sex chromosomes compared to the autosomes (the Y and W have ¼ of the effective population size of autosomes), and result in gradual loss of genetic diversity including beneficial genetic diversity (Muller 1964). The other two evolutionary forces increase deleterious effects. Genetic hitchhiking is driven by selection for a strongly beneficial allele, resulting in linked mildly deleterious variation also increasing in frequency (they hitchhike), since there is no recombination to disassociate it from the beneficial allele. In other words, lack of recombination leads the best haplotype to increase in frequency even if it contains mildly deleterious alleles (Rice 1987b). Background selection is driven by selection against a strongly deleterious allele, which simultaneously acts against linked mildly beneficial genetic variation (Charlesworth 1991; Charlesworth 1996).

The contribution of recombination suppression between sex chromosomes to degeneration has clearly been demonstrated in species across the tree of life in the form of gene loss, accumulation of transposable elements, and reduction of gene expression of the Y- and W-linked alleles (Ohno 2003; Bachtrog 2005; Bachtrog 2006; Zhou et al. 2014). Highly differentiated sex chromosomes typically consist of distinct chromosomal locations affected by various extents of degeneration, which is consistent with recombination suppression happening in chromosomal blocks at different evolutionary times in a stepwise manner. These blocks are termed evolutionary strata, and have distinct levels of genetic differentiation. Some studies have questioned the existence of sharp boundaries between strata, suggesting that each stratum could be formed gradually and the identified sharp boundaries are consequences of the data and methodology used to identify the strata (Bergero et al. 2007; Xu et al. 2019; Furman et al. 2020).

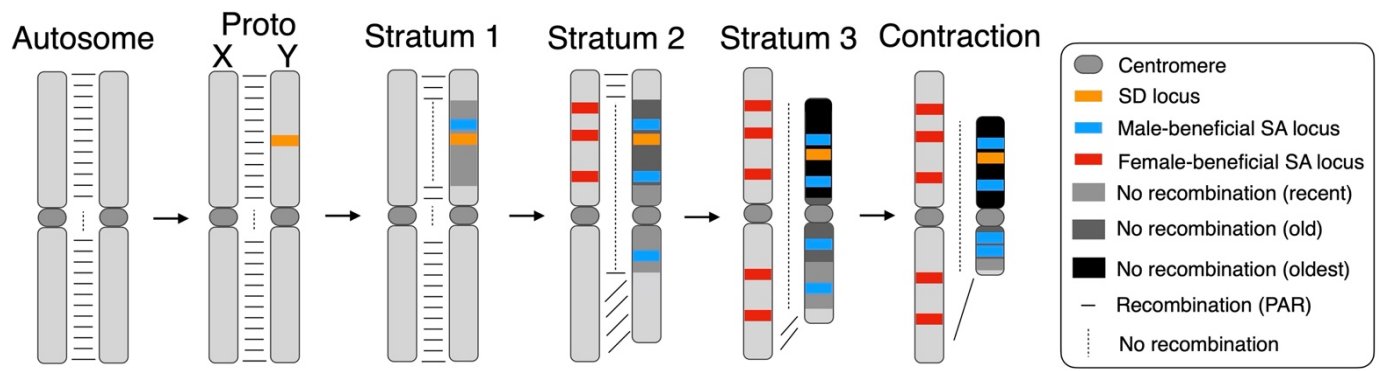

**Figure 2**. Illustration of the canonical model of sex chromosome evolution for an XY system (similar principles apply in a ZW system). A new sex-determining (SD) locus arises on an autosome and the presence of a male-beneficial locus selects for recombination restriction between the two. Additional male- and female-beneficial loci select for an expansion of area of restricted recombination. Regions with no recombination are associated with degeneration such as gene loss, the accumulation of repeats including transposable elements and formation of heterochromatin. Eventually the non-recombining area loses genetic material and shrinks, and only a small recombining area remains, the pseudo-autosomal region (PAR). The figure is adapted from Beukeboom & Perrin (2014).

The canonical model has been widely accepted as the most plausible explanation of sex chromosome evolution (Figure 2). Its predicted stepwise degeneration is observed in highly differentiated and degenerated sex-limited chromosomes in model organisms including mammals, *Drosophila* and most birds (Charlesworth 1991; Ohno 2003; Bachtrog et al. 2011). For example, the human Y chromosome was found to have four distinct evolutionary strata, the oldest one being over 200 Million years old (Lahn and Page 1999; Pandey et al. 2013). Similarly, *Drosophila* and birds have lost the majority of their functional genes and are highly enriched in repeats and heterochromatic regions on their Y and W chromosomes, respectively (Handley et al. 2004; Carvalho et al. 2009; Zhou et al. 2014; Bachtrog et al. 2019).

# Little empirical support for SA selection

Advancements in sequencing technology, methodology, and software development have allowed the sequencing and detailed study of sex chromosomes of many non-model organisms (Bachtrog et al. 2014; Beukeboom and Perrin 2014; Muyle et al. 2017; Vicoso 2019; Kratochvíl et al. 2021; Ma and Rovatsos 2022). This has revealed a remarkable diversity in the rate of sex chromosome differentiation and dynamics. There has been little progress towards empirical support for the role of SA genes in driving sex chromosome recombination suppression and ensuing degeneration (Ironside 2010; Charlesworth 2017; Kratochvíl et al. 2021; Perrin 2021). The issue is that SA gene identification and demonstration of their phenotypic effects is difficult and only possible for sex chromosomes that are still at an early stage in their evolution and maintain SA variation. In addition, demonstrating causality of SA variation for recombination suppression requires closely related species with both recombination and recombination arrest.

Many cases of sex chromosome recombination arrest in the absence of SA genes have been reported, which challenge the universal role of SA genes in driving it (Jeffries et al. 2018; Ma et al. 2018a; Ma et al. 2018b; Veltsos et al. 2020; Perrin 2021). For instance, in the common European frog *Rana temporaria*, there is (homomorphic) sex chromosome differentiation without evidence of SA selection in gene expression and phenotypic data from populations with different levels of sex chromosome differentiation (Ma et al. 2018a; Ma et al. 2018b; Perrin 2021). In fact, recombination suppression between X and Y chromosomes is due to extremely biased recombination in the telomeres of all chromosomes in males (females recombine across the length of their chromosomes) (Berset-Brändli et al. 2008; Brelsford et al. 2016; Furman and Evans 2018; Jeffries et al. 2018). Such extreme heterochiasmy with male-specific reduced recombination is common in many amphibians and fishes (Bergero et al. 2019; Sardell and Kirkpatrick 2019; Cooney et al. 2021; Peterson and Payseur 2021). Functionally, it is similar to complete lack of recombination (achiasmy), as in XY males in *Drosophila* and in ZW females in butterflies and moths (Rastas et al. 2013; John et al. 2016; Satomura et al. 2019). SA selection is also unlikely to be involved in recombination arrest in the anther-smut fungus (*Microbotryum*) which has isogamy. While there are no differences between *Microbotryum* mating types, there is stepwise degeneration of the chromosomes containing the mating-type locus, similar to the evolutionary strata of mammals (Branco et al. 2017; Bazzicalupo et al. 2019; Ma et al. 2020). Studies on a variety of stages of sex chromosome evolution from homomorphic to highly degenerated states are still needed to fully understand why and how sex chromosomes stop recombining and degenerate in some lineages but not in others.

# Sex chromosome age and differentiation do not always reliably correlate

The canonical model of sex chromosome evolution suggests irreversible degeneration and a correlation between divergence time and genetic differentiation between the two sex chromosomes (Figure 2) (Charlesworth and Crow 1978; Charlesworth 1991; Ohno 2003). This is consistent with the highly degenerated and old Y chromosomes in all mammals (>200Mya) and *Drosophila*, and W chromosomes in most birds (> 150Mya) (Presgraves 2008; Cortez et al. 2014; Zhou et al. 2014). However, many other lineages across the tree of life do not follow the degeneration path predicted by the canonical model of sex chromosome evolution. There is strong heterogeneity in the extent of sex chromosome degeneration across lineages, even among closely related species (Stöck et al. 2011; Jeffries et al. 2018; Rovatsos et al. 2019; Xu et al. 2019). Sex chromosome age does not correlate well with the degree of recombination suppression, as sex chromosomes can remain homomorphic or maintain a large recombining area (the pseudoautosomal region) despite long divergence times (e.g. 25-150 million years in ratites) (Stöck et al. 2011; Yazdi and Ellegren 2014; Gamble et al. 2015; Jeffries et al. 2018; Kostmann et al. 2021). Degeneration does not seem to be the inevitable outcome of sex chromosome evolution.

Three possible reasons have been suggested that restrict sex chromosome degeneration: i) haploid selection, ii) sex chromosome turnover and iii) sex chromosome recombination in sex-reversed individuals (the fountain of youth hypothesis; see details below). In organisms where the haploid phase offers considerable opportunities for selection, for example when many genes are expressed, such as in flowering plants, the Y or W chromosomes degenerate slower (Crowson et al. 2017; Scott and Otto 2017; Krasovec et al. 2018). Frequent sex chromosome turnover (see details below) and/or sex chromosome recombination in XY sex-reversed females can explain the old yet homomorphic sex chromosomes in frogs (Stöck et al. 2011; Dufresnes et al. 2015; Jeffries et al. 2018; Rodrigues et al. 2018). However, none of these mechanisms could explain the limited degeneration of the sex chromosomes in ratites, and the study of sex chromosome recombination and genetic structure among closely related species in a phylogenetic framework might offer an alternative explanation.

## Sex chromosome turnover

Unlike the highly stable and well conserved sex chromosomes in mammals and birds, there is frequent birth and death of sex chromosomes, also known as sex chromosome turnover, in many non-model organisms (Figure 3). It has been documented in amphibians, non-avian reptiles, fishes, and certain insects (Dufresnes et al. 2015; Gamble et al. 2015; Vicoso and Bachtrog 2015; Jeffries et al. 2018; El Taher et al. 2021). The reasons for turnovers are one of the big unsolved questions in sex chromosome evolution. Sex chromosome turnover events can be categorised as i) *trans*-heterogamety transition, from an XY system to a ZW system or vice versa; ii) heterologous *cis*-heterogamety transition, which retains the same heterogametic sex (XY to XY or ZW to ZW); iii) homologous *cis*-heterogamety transition, which changes the sex-determining gene on the same Y or W chromosome (Figure 3) (Saunders 2019; Furman et al. 2020).

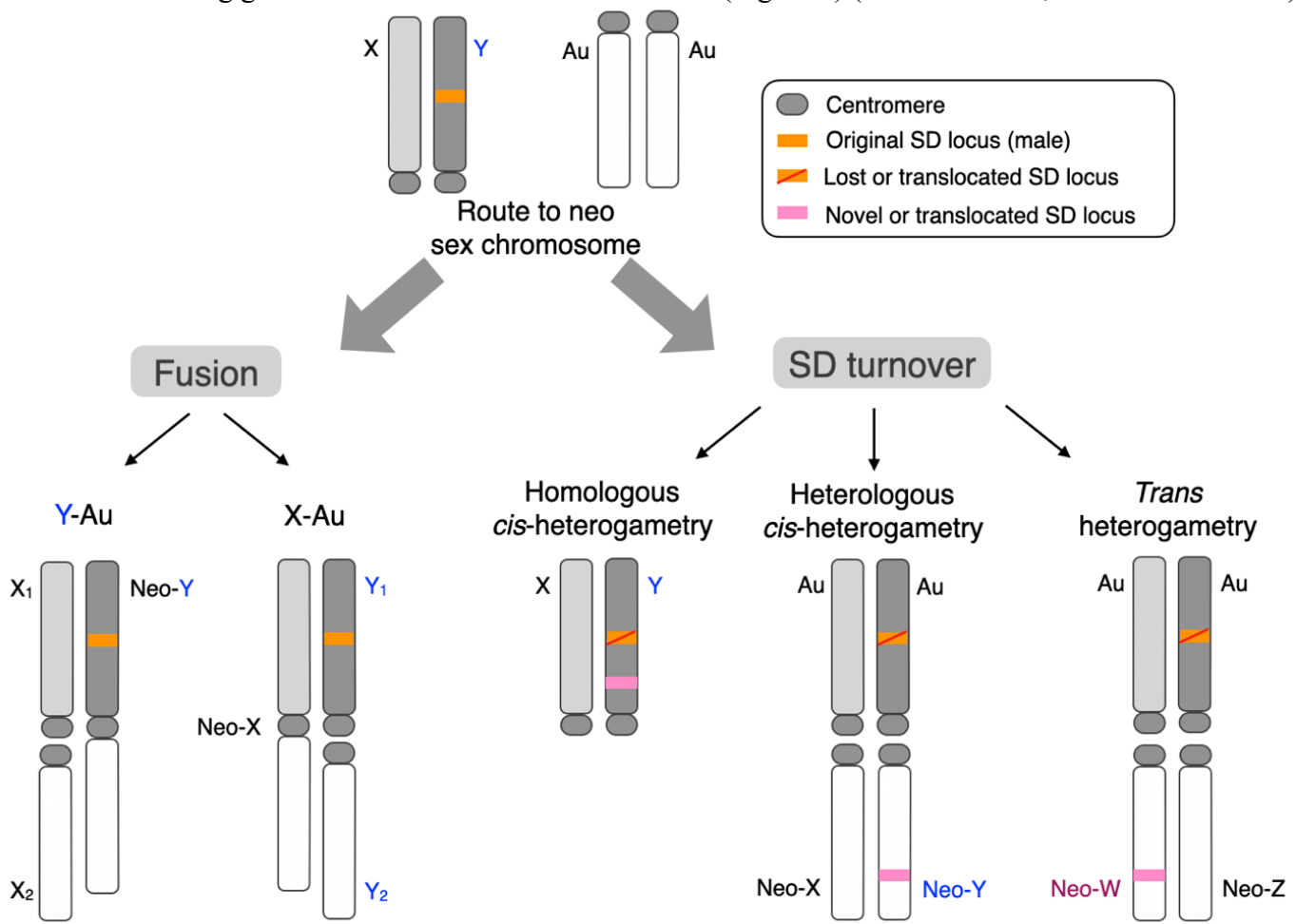

**Figure 3**. Illustration of sex chromosome evolution routes to form neo/new sex chromosomes. One involves chromosomal fusions, and the other involves sex chromosome turnover. Turnover can be categorized into: homologous *cis*-heterogamety where the original sex determining gene (yellow) is replaced or translocated to a new genomic position in the same chromosome (pink); heterologous *cis*-heterogamety, where the original SD gene (yellow) is replaced or translocates to a different chromosome (pink); and *trans*-heterogrametry, where the original SD gene (yellow) is replaced or translocates to a different chromosome and the heterogametic sex inverts. The ancestral sex chromosomes (grey) could be from any stage of the canonical model.

Sex chromosome turnover can occur: i) if the existing sex-determining region is translocated to an autosome, possibly facilitated by transposable elements; or ii) if a new gene overtakes the sex-determining role via mutation (Kitano and Peichel 2012; Jeffries et al. 2018; El Taher et al. 2021; Kabir et al. 2022). Sex chromosome turnover is usually studied by identifying which chromosome has sex-linked markers. This allows to rapidly sample many lineages, but does not uncover the underlying genetic mechanism (Rovatsos et al.; Gamble et al. 2015; Jeffries et al. 2018; Keating et al. 2022). Furthermore, it cannot detect sex-determination transitions within the same sex chromosome (homologous *cis*-heterogeneity), thus underestimating the true turnover rate. To elucidate the genetic cause of sex chromosome turnover, it is necessary to compare the identity and location of the sex-determining gene in closely related species. This approach has revealed repeated translocation of a sex-determining region to be responsible for sex chromosome turnover in salmonids and *Takifugu* fishes (Phillips 2013; Kabir et al. 2022). The alternative possibility of a new mutant sex-determining locus replacing the existing one has been suggested, but not conclusively demonstrated, by studies finding different autosomes used as the sex chromosome in various closely related species. The new sex chromosomes are a non-random subset of all possible chromosomes enriched in candidate genes for sex determination (Jeffries et al. 2018); however, there are neither comparative genomic studies to identify sex-determining regions, nor functional studies that demonstrate those genes to be responsible for sex determination.

Computer simulations and analytical studies have proposed four potential drivers of sex chromosome turnover which can act separately or in combination (Saunders 2019). They are genetic drift (Veller et al. 2017; Saunders et al. 2018), deleterious mutation load (Blaser et al. 2012; Blaser et al. 2014), sex-ratio selection, which includes genomic conflicts involving biased sex chromosome transmission such as meiotic drive (Kozielska et al. 2006; Cordaux et al. 2011) and SA selection (Van Doorn and Kirkpatrick 2007; Van Doorn and Kirkpatrick 2010; van Doorn 2014). In the neutral model, an emergent sex-determining gene simply drifts to fixation. The predicted turnover is 2-4 times more likely for *cis* compared to *trans* heterogametry (Saunders et al. 2018). Under the deleterious mutation load hypothesis (the "hot potato" model), deleterious mutations that accumulate on the ancestral sex chromosomes increasingly favour *cis*-heterogametry turnovers in which the old mutation-loaded Y or W chromosome is lost and replaced by a new, mutation-free one. *Trans*-heterogametry transitions are unlikely as they would involve the fixation of the mutation-loaded Y or W (Blaser et al. 2012). The sex-ratio selection hypothesis involves the biased transmission of sex chromosomes, leading to a biased sex ratio which can be selectively advantageous. For instance, a Y-chromosome meiotic driver results in a male-biased sex ratio. This creates a strong selective pressure for a dominant female sex-determining gene on the X chromosome or an autosome, which turns some genetic males into females, and can lead to a transition from male to female heterogamety (Kozielska et al. 2010). Finally, the SA selection hypothesis suggests that the fitness benefits of genetic linkage of a new sex-determining gene with a SA gene could cause a sex chromosome turnover (Van Doorn and Kirkpatrick 2007). These sex chromosome transitions are as likely to be caused by masculinising as they are by feminising sex-determining mutations, so that an equal transition rate between XY and ZW systems is predicted.

The contribution of the above forces to sex chromosome turnover can be studied with a comparative genomic approach, as in the large vertebrate dataset from the tree of sex project (Ashman et al. 2014; The Tree of Sex Initiative v2). It clearly showed clade-specific patterns in turnover rates between ZW and XY systems. The transition rate from ZW to XY is significantly higher than the reverse in fish, while there was no such difference in squamates or amphibians (Pennell et al. 2018). Transitions to XY are favoured by stronger sexual selection in males than females; however, it is unclear why this would only apply to fish (Pennell et al. 2018) (Bachtrog et al. 2011; Rice, 1986). Another explanation is dominant masculinizing mutations that generate neo-Y chromosomes, which protect against the female-biased sex ratio caused by cytoplasmic sex ratio distorters (Beukeboom & Perrin, 2014). Again, it is unclear if fishes have more such distorters than amphibians or squamates.

Sex chromosome turnover, together with the 'fountain of youth' theory (see details in next section), were the hypothesized major mechanisms to maintain homomorphic sex chromosomes in the Ranidea true frogs and *Hyla* tree frogs (Dufresnes et al. 2015; Jeffries et al. 2018). However, other amphibians, fishes, and reptiles with heteromorphic sex chromosomes have also had sex chromosome turnovers (Furman et al. 2020; Ma and Veltsos 2021). Sex chromosome turnover therefore does not seem to be a general mechanism to maintain homomorphic sex chromosomes, as it applies to specific clades even within frogs.

# Fountain of youth

The "fountain of youth" theory states that deleterious mutations could be purged from the sex-limited chromosomes (Y or W) and long-term differentiation could be prevented by occasional sex-chromosome recombination in sex reserved individuals (i.e. XY females) (Perrin 2009). Multiple lines of supporting evidence have been found in European tree frogs. Their sex chromosomes remained homomorphic despite the lack of XY recombination in males, and haplotypes at sex-linked markers further cluster by species and not by gametologs, suggesting a history of recurrent XY recombination (Stöck et al. 2011; Dufresnes et al. 2015). Using an Approximate Bayesian Computation approach, Guerrero et al. (2012) showed the rate of XY recombination in this tree frog group was significantly different from zero, while being $\sim 10^{-6}$ lower than that of XX recombination. A possible mechanism allowing XY recombination is the occasional sex reversal of XY individuals (i.e., sex-reversed XY females), resulting from incomplete genetic control over sex determination (Perrin 2009). If the arrest of XY recombination in males is a property of male meiosis and not the male genotype, then Y chromosomes would recombine with X chromosomes in the occasional sex-reversed XY females, preventing Y chromosomes from progressive differentiation and degeneration. Sex-reversal experiments in a series of taxa (crested newts, medaka fish, housefly) have confirmed that sex differences in recombination are largely depend on phenotypic, not genotypic, sex (Inoue et al. 1983; Kondo et al. 2001; Rodrigues et al. 2018). Additional support for the fountain-of-youth model comes from genetic mapping in wild populations of the common frog *R. temporaria*, which showed that XY recombination only depends on phenotypic sex. Wild XX males showed recombination restriction similar to XY males, while wild sex-reversed XY females recombined as much as XX females (Rodrigues et al. 2018).

# Expanded Y or W chromosomes and gene gains

The small size and loss of genes on the Y or W chromosomes of mammals, birds, and many insects are considered a hallmark of sex chromosome differentiation and degeneration (Bachtrog 2006; Bachtrog 2008; Wright et al. 2016). However, the expansion of Y/W chromosomes has also been observed in lineages of fishes, amphibians, and flowering plants, which was not specifically addressed as a necessary evolutionary step in the canonical model (Fisher 1931; Rice 1987a; Charlesworth 1996; Charlesworth et al. 2005). The expansion of Y and W chromosomes can be a rapid and effective mechanism to differentiate them from their meiotic partners and further prevent crossovers between sex chromosomes. The shrinking of the Y and W chromosomes to their familiar small size in mammals and birds is proposed to occur at a later phase of their degeneration (Ming et al. 2011; Schartl et al. 2016). The giant Y/W chromosomes of some species are thus considered 'younger' than the shrunk Y/W chromosomes of other species.

It has been suggested that repetitive sequences that accumulated at the early stage of sex chromosome evolution are often responsible for the size expansion, which could ultimately facilitate the arrest of recombination between the sex chromosomes (Ming et al. 2011; Schartl et al. 2016). Data from plants also identify enrichment of particular repeats on the sex chromosomes, but can be on either the X or Y (Hobza et al.; Cermak et al. 2008; Kubat et al. 2008; Kralova et al. 2014). One study in fish suggests an association between the amount of repetitive sequence accumulation and the age of early-stage sex chromosomes (Chalopin et al. 2015), suggesting a role of TE accumulation on sex chromosome differentiation. However, it did not take into account possible changes in coding sequence, and the genome assemblies were too fragmented to assess the possibility of gene gain. In sharp contrast to gene loss as the hallmark of sex chromosome degeneration, in *Drosophila*, the rate of gene gain on the Y is greater than gene loss (Carvalho et al. 2009) and there can also be extensive gene amplification via tandem duplication as in *Drosophila miranda* (Bachtrog et al. 2019). More studies on detailed gene content of early-stage sex chromosomes across taxa would help address the prevalence of massive gene amplification on the sex-limited chromosomes (or mating-type chromosomes) across taxa. Together, they will ultimately extend the existing theory of the canonical model of sex chromosome evolution.

Regardless of gene gain or loss, the result is an unbalanced gene copy number between the sex chromosomes. This is alleviated by the evolution of dosage compensation, which prevents possible detrimental effects by balancing the gene expression between the two sexes (Ohno 1966; Mank 2013).

Mechanisms of sex chromosome dosage compensation differ strikingly among animals, and have been extensively reviewed previously (Mank 2009; Harrison et al. 2012; Mank 2013; Furman et al. 2020).

Another way for generating large sex-limited chromosomes is a fusion of the W or more often of the Y (as has happened many times in placental mammals, iguanas and teleost fish) with an autosome (Pennell et al. 2015), producing multiple neo-sex chromosomes, where the newly added parts behave as pseudoautosomal regions and can go through subsequent differentiation. Recent studies in vertebrates demonstrated that XY and ZW sex chromosome systems differ in the frequency of fusion between the sex chromosomes with an autosome. For instance, multiple neo-sex chromosomes evolved independently more than 20 times in mammals and around 15 times in iguanas with XY systems, whereas only 6 times in caenophidian snakes and maybe just once in birds with a ZW system (Pennell et al. 2015). This may reflect faster differentiation of the Y in comparison to the W owing to higher mutation bias or stronger selection in males (Ohno 2003; Furman et al. 2020). Alternatively, the differential involvement of sex chromosomes in female meiosis, particularly in female meiotic drive (Smith et al. 2016) may be involved in this bias. This could be explained by sex ratio selection driving the turnover of sex chromosomes (see details in 'Sex chromosome turnover' section).

## Alternative models of sex chromosome evolution

Recombination suppression is universally associated with sex chromosome evolution (Olito and Abbott 2023). Recombination rate can be affected by gene density, the presence and type of repetitive elements, chromosomal rearrangements, and chromatin structure, and can be sexually dimorphic (Broman et al. 1998; Stevison et al. 2011; Kent et al. 2017). These factors can change neutrally or through selection and affect the evolution of sex chromosomes. It is crucial to understand the evolutionary forces and mechanisms responsible for recombination suppression to understand sex chromosome degeneration. Given the lack of evidence for SA selection to drive recombination suppression, several alternative theories that do not rely on the presence of SA genes have been proposed (Figure 4).

### Pre-existing recombination suppression between sex chromosomes

It is possible that sex-determining genes evolve in regions of pre-existing recombination suppression (Ortiz-Barrientos et al. 2016; Bergero et al. 2019; Charlesworth 2019). Of particular interest are cases of sex-specific recombination, which has been observed in various lineages, ranging from significant reduction to complete absence in one sex (Figure 4A). In achiasmy there is no recombination in all chromosomes in one sex, as in XY males in *Drosophila* and ZW females in butterflies and moths (Figure 4A1) (Lenormand 2003; Lenormand and Dutheil 2005; Satomura and Tamura 2016; Satomura et al. 2019; Sardell and Kirkpatrick 2020). In extreme heterochiasmy, recombination is concentrated on chromosome ends in one sex while it is spread over the whole chromosome length on the other (Figure 4A2). The outcome is a drastic recombination reduction over most of the genome in one sex. This is the case in guppies and frogs (Berset-Brändli et al. 2008; Brelsford et al. 2016; Jeffries et al. 2018; Bergero et al. 2019).

Sexually dimorphic recombination could be adaptive. The few reproducing individuals of the sex with the highest variance in reproductive success, typically the males, might benefit from lack of recombination if their ability to reproduce was favorably influenced by their genetic makeup (Haldane 1922; Huxley 1928). However, this would make achiasmy in moths and butterflies maladaptive. The rapid rate of turnover in achiasmy in closely related species suggests it could evolve to adaptively restrict recombination on sex chromosomes (Satomura et al. 2019). Similarly, extreme heterogametry could adaptively evolve to affect the heterogametic sex, however, to date mostly male extreme heterochiasmy has been reported in fishes and frogs. The one exception is the Kajika frog (*Buergeria buergeri*) with a homomorphic ZW system, which shows a largely ring-shaped ZW bivalent during female meiosis with mostly zero or one chiasmata, while the autosomes showed recombination across their length, suggesting restricted recombination has evolved in a ZW system at least once (Ohta 1986). Alternatively, pre-existing extreme heterochiasmy might be due to mechanistic constrains such as differences between male and female meiosis (Sardell and Kirkpatrick 2020).

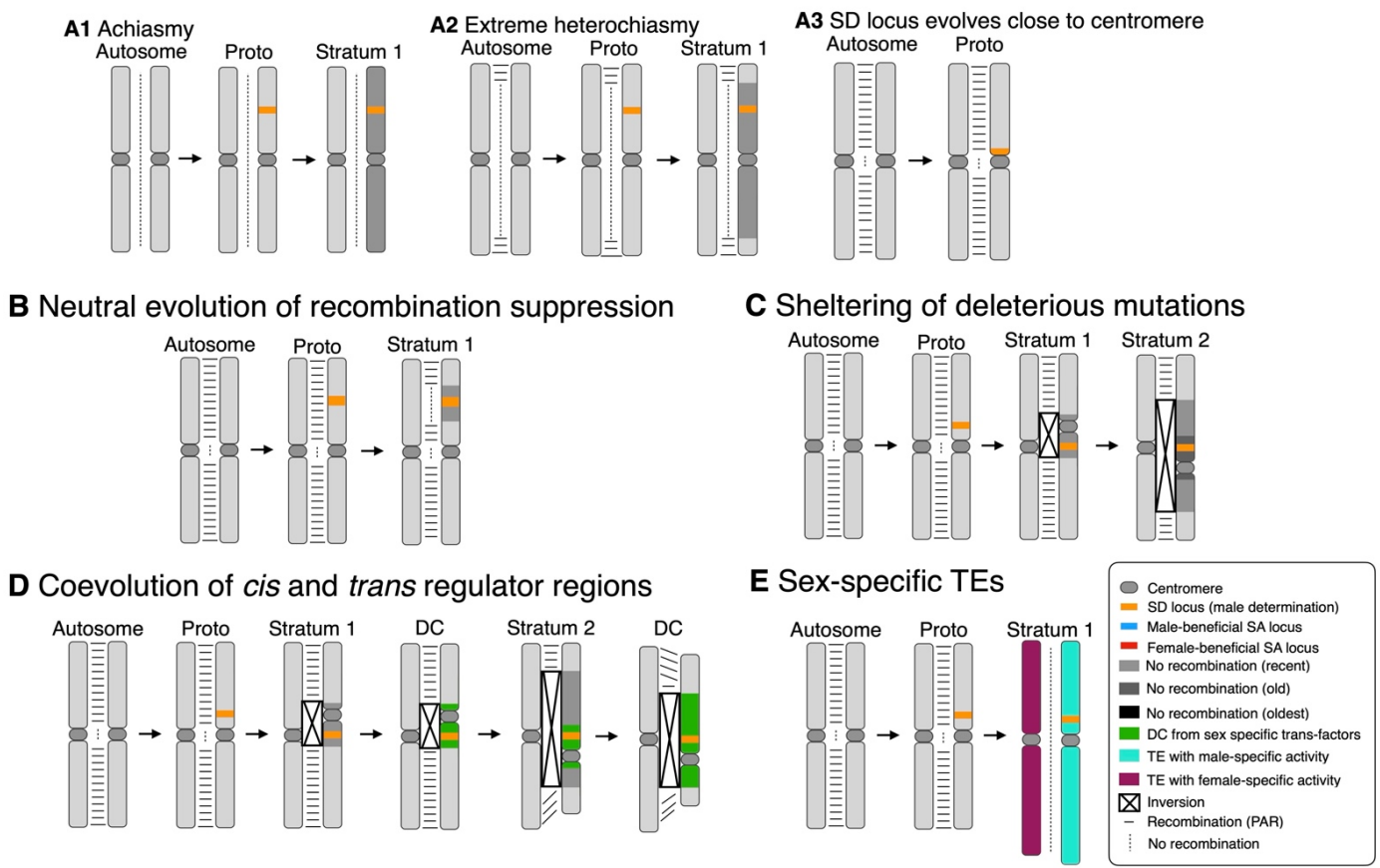

**Figure 4**. Main alternative theories to drive recombination suppression between sex chromosomes without invoking SA selection. A. Pre-existing recombination suppression. A1. Sex-specific achiasmy, A2. Extreme heterochiasmy. Both result in immediate recombination cessation for the whole chromosome where the SD locus evolves. A3 Evolution of the SD locus in region of pre-existing low recombination, such as the centromere. B. Neutral fixation of modifiers that reduce recombination close to the SD locus. C. Sheltering of deleterious mutations by a lucky inversion that captures a haplotype with fewer deleterious mutations than average and the SD gene. The process can repeat to give rise to evolutionary strata. D. Starts as C but *cis* degeneration of genes close to the SD locus result in dosage compensation, that prevents future reversion of the inversion. E. Male- (turquoise) and female- (purple) specific TEs, increase in frequency on the proto-sex chromosomes, causing loss of homology and reduced recombination.

Several studies have identified sex-determining genes in regions with pre-existing low recombination. There are many cases in XY frogs, in which new sex-determining genes immediately stop recombination over the whole Y length (Jeffries et al. 2018; Perrin 2021). The sex-determining locus in guppies was found in an existing non-recombining region (Bergero et al. 2019; Charlesworth 2019). There are also plant examples such as kiwi and papaya where the identified sex-determining region is close to the centromere which has low recombination (Figure 4A3) (Iovene et al. 2015; Tiley and Burleigh 2015; Pilkington et al. 2019; Rifkin et al. 2022).

Pre-existing sexual dimorphism in recombination can account for the presence of low recombination in a newly formed sex-determining region without the need to invoke SA genes. However, it cannot account for stepwise sex chromosome degeneration, which requires additional recombination suppression events. Studies of the evolutionary forces and/or genetic mechanism behind extreme heterochiasmy are needed to better understand how they contribute to sex chromosome evolution and stepwise degeneration.

**Neutral evolution of recombination suppression**

It is possible for recombination suppression on the sex chromosomes to evolve under neutral conditions, based on the principle that their heterozygosity decreases the probability of recombination (Ironside 2010). It is also possible for neutral accumulation of sequence divergence adjacent to the sex-determining locus to drive recombination suppression between sex chromosomes and lead to differentiation (Jeffries et al. 2021). Computer simulations show that with high mutation rate and small population size and recombination rate, a region with restricted recombination can emerge and spread outward from the locus responsible for sex determination, purely driven by neutral processes (Bengtsson and Goodfellow 1987; Eliis et al. 1990; Jeffries et al. 2021). Theoretically, this process should apply to any genomic region that remains strictly heterozygous, such as around mating-type loci, supergenes found in certain ant species on social chromosomes, or loci governing dimorphisms like distyly in plants (Schwander et al. 2014; Branco et al. 2017; Brennan 2017; Martinez-Ruiz et al. 2020). Consequently, neutral divergence accumulating in linkage disequilibrium with the sex-determining locus decreases recombination rate locally, establishing a positive feedback loop that promotes further divergence (Jeffries et al. 2021). The limitation of this model is that biologically realistic cases of appropriately small population size and high mutation rate are challenging to find (Jeffries et al. 2021). While the model is theoretically reasonable, it is rather unpractical to test empirically.

**Sheltering of deleterious mutations in heterozygous regions on sex chromosomes**

Another consequence of the permanent heterozygosity of the sex chromosomes in the heterogametic sex is that inversions on the Y or W chromosomes that capture haplotypes with fewer deleterious mutations than average will have a fitness advantage and increase in frequency (Jay et al. 2022). Additional lucky inversions that also capture a haplotype with lower deleterious load than average can successively accumulate on the Y and W chromosomes following the same principle. The outcome is the formation of evolutionary strata, i.e. regions that stopped recombining at different times. The sheltering of deleterious mutations model does not work well in species with large effective population sizes or low mutation rates, and can explain the prevalence of homomorphic sex chromosomes in such species (Jay et al. 2022). The model also applies to other permanently heterozygous genomic regions, such as supergenes housing more than two permanently heterozygous alleles as in butterflies and ants, plant self-incompatibility loci, mating-type loci of dikaryotic fungi and the association between meiotic drivers, often permanently heterozygous, and large non-recombining regions involving polymorphic chromosomal inversions (Dyer et al. 2007; Reinhardt et al. 2014; Jay et al. 2022; Yan et al.; Branco et al. 2017; Hartmann et al. 2021; Jay et al. 2022). Importantly, it can account for the extension of the non-recombining region without invoking SA loci.

The sheltering theory has been criticized. Using the same simulations, another study showed that even if the initial inversion captures a region of low genetic load, the lack of recombination will cause it to quickly accumulate more deleterious mutations than alternative haplotypes. This will either lead to its population extinction, or cause very strong selection to re-invert the region and restore recombination (Lenormand and Roze 2024). Another study showed that the initial sheltering simulation study estimated fixation probabilities of Y-linked inversions after excluding inversions that were lost in the first 20 generations (Olito and Charlesworth 2023). Their reanalysis states that if all simulation runs are included, most simulated parameters yield estimates of fixation probabilities that are close to the neutral expectation. Therefore, the proposed sheltering mechanism is unlikely to provide a robust selective advantage to inversions suppressing recombination between evolving sex chromosomes. Both studies suggest that, for a brief period after its evolution, the lucky inversion can rise in frequency, but additional evolutionary forces are required to ensure its long-term survival.

**Coevolution of *cis* and *trans*-regulatory regions contribute to stepwise sex chromosome degeneration**

The coevolution of *cis* and *trans* regulators of gene expression for the chromosome restricted to the heterogametic sex plays a role in recombination suppression between sex chromosomes and further the stepwise sex chromosome degeneration. It can maintain recombination suppression even when deleterious effects accumulate (Lenormand et al. 2020; Lenormand and Roze 2022). This model starts with the fixation of a lucky inversion similarly as in the sheltering model. In the initial inversion region, regulatory regions on X and Y chromosomes begin independent evolution and accumulate divergence. This initiates a rapid degeneration of Y-linked alleles, and regulatory changes associated with dosage compensation evolve, which further contribute to sex chromosome differentiation and degeneration. The resulting dosage compensation hinders the possibility of re-establishing recombination in the long run. The impact of Y recombination arrest and degeneration is expected to occur more rapidly in larger populations. No sexually antagonistic loci are required, the sexual antagonism is an emergent consequence of the interaction between the *cis* and *trans* loci that control dosage compensation.

### Sex-specific transposable element activity

Transposable elements and other repeats are often enriched on sex chromosomes (Cermak et al. 2008; Kralova et al. 2014). Their accumulation on sex chromosomes has been interpreted as one of the consequences of loss of recombination in the canonical model (Charlesworth 1996; Bachtrog 2005). An alternative explanation is that the TEs accumulate because they have sex-specific activity as has been demonstrated in plants (Hobza et al. 2015; Hobza et al. 2017; Kent et al. 2017). This raises the possibility that the accumulation of repeats comes first and recombination suppression is a consequence, either of the sequence divergence caused by the presence and absence of repeats, or by the additional regulatory mechanisms that respond to high repeat localization, such as heterochromatinization (Hobza et al. 2015; Hobza et al. 2017; Muyle et al. 2021). The study of sex-specific activity of TEs in meiotic tissue should be expanded to more organisms to evaluate whether TE accumulation precedes rather than supersedes recombination suppression. If it turns out to be the case, sex-specific TE activity could be a major force in sex chromosome evolution.

## Conclusion and future directions

The canonical model of sex chromosome evolution is not universally supported. Many lineages experience frequent sex chromosome turnovers or maintain homomorphic sex chromosomes over long periods of time, suggesting degeneration is not always a matter of time. The sex-limited Y/W chromosomes can expand before they contract in size. Both TEs and gene gains could contribute to this size expansion, which further challenges gene loss being the hallmark of sex chromosome degeneration. Empirical support for a role of SA selection as a driver of recombination suppression on sex chromosomes remains elusive and multiple models that result in loss of recombination without invoking SA selection have been proposed, although they have not been empirically studied yet.

It seems there is no single trajectory for sex chromosome formation and evolution across the tree of life, suggesting the underlying mechanisms and evolutionary forces are diverse and lineage specific. The most informative systems for future studies may be those exhibiting the most variation in sex chromosome divergence or turnover because the generated comparisons would allow to distinguish cause and effect. Furthermore, studies of early-stage sex chromosomes are crucial because the genetic causes for sex chromosome formation, such as SA variation, are likely to become lost or become too noisy to study in old sex chromosomes.

Future work on the underlying forces and genetic mechanisms driving sex chromosome turnovers is expected to be fruitful. It should consider the importance of ecologically relevant selection (Meisel 2022). Identifying how and what makes the sex-determining mechanisms so labile in certain lineages, but very stable in others, remains a challenge. Finally, alternative theories of sex chromosome evolution should be empirically tested with data from various lineages. The collation of genomic data from diverse taxa across the tree of life such as in the tree of sex initiative (https://www.sanger.ac.uk/project/tree-of-sex-initiative/), promises to be a great resource to further understand how sex chromosomes evolve.

**Author Contributions**: W.-J. M. designed the study. P. V., W.-J. M. and S. S. collected data and synthesized ideas from literature. P. V. and W.-J. M. visualized the synthesized ideas with input from S. S. W.-J.M. and P.V. wrote the manuscript, with significant input from S. S. All authors have read and agreed to the submitted version of the manuscript.

**Data Availability Statement**: NA.

**Acknowledgments**: This work is funded by the European Union (ERC starting grant, FrogWY, 101039501) and a starting grant from Vrije Universiteit Brussel (OZR4049) to Wen-Juan Ma. We thank anonymous